\documentclass[aps,prc,showpacs,showkeys,floatfix,footinbib,twocolumn,
10pt]{revtex4-1}
\usepackage{subfigure}
\usepackage{amsmath}    
\usepackage{graphicx}   
\usepackage{verbatim}  
\usepackage{color}      
\usepackage{hyperref}
\long\def\/*#1*/{}

\usepackage{filecontents}

\RequirePackage{lineno}

\begin{document}


\title{Centrality dependence of chemical freeze-out parameters from
net-proton and net-charge fluctuations using hadron resonance gas model}

\author{Rama Prasad Adak}
\author{Supriya Das}
\author{Sanjay K. Ghosh}
\author{Rajarshi Ray}
\author{Subhasis Samanta}
\email[E-mail: ]{subhasis.samant@gmail.com}
\altaffiliation[Present Address: ]{School of Physical Sciences, National
Institute of Science Education and Research, Jatni-752050, India}
\affiliation{Center for Astroparticle Physics \& Space Science, Bose
Institute, Block-EN, Sector-V, Salt Lake, Kolkata-700091, India 
\\ \& \\ 
Department of Physics, Bose Institute, \\
93/1, A. P. C Road, Kolkata - 700009, India}

\begin{abstract}
 We estimate chemical freeze-out parameters in HRG and EVHRG
model by fitting the experimental information of net-proton
and net-charge fluctuations measured in Au + Au collisions by the STAR
collaboration at RHIC.  We observe that chemical freeze-out parameters
obtained from lower and higher order fluctuations are
though almost same for $\sqrt{s_{NN}} > 27$ GeV, tend to deviate from
each other at lower $\sqrt{s_{NN}}$.  Moreover, these separations
increase with decrease of $\sqrt{s_{NN}}$ and for a fixed
$\sqrt{s_{NN}}$ increase towards central collisions.  Furthermore, we
observe an approximate scaling behaviour of
$(\mu_B/T)/(\mu_B/T)_{central}$ with $(N_{part})/(N_{part})_{central}$
for the parameters estimated from lower order fluctuations
for 11.5 GeV $\le \sqrt{s_{NN}} \le$ 200 GeV.  Scaling is violated for
the parameters estimated from higher order fluctuations for
$\sqrt{s_{NN}}= 11.5$ and 19.6 GeV.  It is observed that the chemical
freeze-out parameter, which can describe $\sigma^2/M$ of net-proton very
well in all energies and centralities, can not describe the $s\sigma$
equally well and vice versa.
\end{abstract}

\pacs{25.75.-q, 24.60.Ky, 24.10.Pa}


\keywords{Heavy Ion collision, Fluctuation, Chemical freeze-out, Hadron
Resonance Gas model}

\maketitle

\section{\label{sec:Intro} Introduction}
Relativistic heavy ion collisions are investigated both theoretically
and experimentally to understand the properties of nuclear matter at
extreme conditions. In heavy ion collisions, there is a possibility for
the nuclear matter to undergo a phase transition to quark matter.  The
nature of the phase transition is still not well established.  At low
baryon chemical potential and high temperature nuclear matter is
expected to smoothly cross over~\cite{Aoki:2006we} to a quark gluon
plasma (QGP) phase. Whereas, at high baryon chemical potential and low
temperature the system is expected to have a first order phase
transition~\cite{Asakawa:1989bq, Ejiri:2008xt, Bowman:2008kc}.
Therefore, the first-order phase transition at high baryon chemical
potential and low temperature should end at a critical end-point (CEP)
as one moves towards a high temperature and low baryon chemical
potential region in the phase diagram of strongly interacting matter
~\cite{Halasz:1998qr, Fodor:2004nz, Gavai:2004sd, Stephanov:2004wx}.
The main goal of experiments of heavy ion collisions is to map the
quantum chromodynamics (QCD) phase diagram in terms of  temperatures and
baryon chemical potentials.  One of the main objectives of the beam
energy scan (BES) program of RHIC is to investigate the location of
CEP. In the near future, CBM experiment at FAIR will also involve in
such an investigation along with the other studies of strongly
interacting matter at high baryon chemical potentials and low
temperatures.

The event-by-event fluctuations of conserved charges like baryon,
strangeness, and electric charge are sensitive indicators of the
transition from hadronic matter to QGP.  Moreover, the existence of the
CEP can be signalled by the divergent fluctuations.  Therefore, a
non-monotonic variation of observables related to the cumulants of the
distributions of the above mentioned conserved charges with a variation
of centre of mass energy ($\sqrt{s_{NN}}$) are believed to be good
signatures of a phase transition and a CEP \cite{Asakawa:2000wh,
Adamczyk:2013dal}. However, this non-monotonic behaviour is a
necessary but not sufficient condition for the CEP.  For example, the
singularities associated with first or second order transition, in the
infinite volume limit, may become finite peaks due to finite volume
effect. Moreover, due to the finite size of the system in heavy ion
collisions, non-monotonic behaviour may be indicative of pseudo-critical
region which is shifted from the actual critical region
\cite{Lacey:2014wqa, Ladrem:2004dw, Palhares:2009tf}. 

It may be expected that with the variation of centrality, keeping
$\sqrt{s_{NN}}$ fixed, similar behaviour as those found for the
variation of centre of mass energy would be observed.  However, the
signatures of phase transition or CEP are detectable only if they
survive during the evolution of the system.  Several experimental
results of conserved charge fluctuations (or cumulants) from BES
program have recently been reported at various energies and
centralities \cite{Adamczyk:2013dal, Adamczyk:2014fia, Adare:2015aqk}.
However, these data do not show non-monotonic behaviour as a function of
$\sqrt{s_{NN}}$.  On the other hand, a new analysis of net proton
moments have been reported by STAR collaboration \cite{Luo:2015ewa}
where the upper $p_T$ coverage for proton and anti-proton has been
extended up to 2 GeV using the time of flight (ToF) detector. In this
analysis a non-monotonic behaviour for higher order cumulants
($\kappa\sigma^2$) at lower $\sqrt{s_{NN}}$ has been reported indicating
a probable CEP like behaviour.   
Finite system size may also cause this non-monotonic
behaviour. In principle such effects may be estimated from the ratio of
cumulants, as discussed in \cite{Bhattacharyya:2015zka} using Hadron
resonance gas (HRG) model for illustration. It has been shown that
though for net proton and net kaon the cumulant ratios are almost volume
independent, the cumulant ratios of net charge are highly sensitive to
the system volume. This is mainly due to the contribution of pions which
are extremely light in the hadronic scale.

Fluctuations which are related to the thermodynamic susceptibilities via
fluctuation-dissipation theorem \cite{Kubo} can be studied using LQCD or
models.  However, since cumulants are volume dependent, ratios of
cumulants are constructed to cancel volume term and they are related to
the ratios of the different order of susceptibilities.  Therefore, it is
possible to extract chemical freeze-out parameters like 
temperature and chemical potential by comparing
experimentally measured ratios of cumulants with ratios of
susceptibilities calculated in LQCD or in a model
\cite{Borsanyi:2014ewa, Alba:2014eba}.  Thus the ratios of cumulants of
conserved charges provide important information about chemical
freeze-out parameters which is useful to locate CEP in the phase
diagram.  However, at finite chemical potential, LQCD faces the
well-known sign problem. As a result, the region of very high chemical
potential in the phase diagram can not be studied in LQCD presently.
Moreover, it is not possible to employ experimental acceptance cuts in
LQCD calculation.  On the other hand, hadron resonance gas (HRG) model
\cite{Hagedorn:1980kb, Rischke:1991ke, Cleymans:1992jz,
BraunMunzinger:1994xr, Cleymans:1996cd, Yen:1997rv,
BraunMunzinger:1999qy, Cleymans:1999st, BraunMunzinger:2001ip,
BraunMunzinger:2003zd, Karsch:2003zq, Tawfik:2004sw, Becattini:2005xt,
Andronic:2005yp, Andronic:2008gu,Begun:2012rf, Andronic:2012ut,
Tiwari:2011km, Fu:2013gga, Tawfik:2013eua, Garg:2013ata,
Bhattacharyya:2013oya, Bhattacharyya:2015zka,Kadam:2015xsa,
Kadam:2015fza, Kadam:2015dda, Albright:2014gva, Albright:2015uua,
Begun:2016cva} provides us with a simpler model for the study of the
strongly interacting matter in the non-perturbative domain.
HRG model is based on the assumption of thermal equilibrium of
a system composed of free hadrons and resonances. One may estimate the
commensurate chemical freeze-out parameters by fitting the experimental
data of various hadronic observables with the HRG model
\cite{Cleymans:2005xv,Xu:2001zj,Becattini:2005xt,Andronic:2005yp,
Andronic:2009jd, Karsch:2010ck,Chatterjee:2015fua}. Also the
susceptibilities of conserved charges calculated in LQCD have been well
reproduced by HRG model \cite{Karsch:2003zq, Tawfik:2004sw,
Andronic:2012ut, Bhattacharyya:2013oya} for temperatures up to 150 MeV.
Moreover the region of large chemical potential in the phase diagram,
which can be accessed by low energy heavy ion collisions, can be studied
by this model. Since, one can incorporate proper experimental
acceptances in this model, it can be used to estimate chemical
freeze-out parameters by fitting experimental data of the ratios of
cumulants of conserved charges. It should be noted however that the final
parameters are still model dependent.

Here we would like to emphasise the salient feature of our
present study. If the system becomes thermalised well ahead of
freeze-out, then all the observables would carry the signature of
thermalisation. In such a scenario the observed hadrons should have a
clear thermodynamic equilibrium distribution. Therefore a thermal model
like HRG would show a very good agreement with the data up to all orders.
Any difference from this scenario may point towards a more complex
system and our attempt here is to find such discrepancies in the
parametrisation of the HRG model from various experimental data and
gain some insight about the system.

The paper is organised as follows. The ideal and excluded volume hadron
resonance gas model are introduced in Sec. \ref{sec:HRG}. In Sec.
\ref{sec:Fluctuations} we have briefly discussed fluctuations of
conserved charges and several relevant experimental observables.  Then
in Sec. \ref{sec:Results} we have discussed results of this paper.
Finally, we summarise our results in Sec. \ref{sec:Discussion}.

\section{\label{sec:HRG} Hadron Resonance Gas model}

In HRG model, the system of thermal fireball consists of all the hadrons
and resonances given in the particle data book \cite{Agashe:2014kda}.
There are varieties of
HRG models in the literature.  Different versions of this model and some
of the recent works using these models may be found in Refs
\cite{Hagedorn:1980kb, Rischke:1991ke, Cleymans:1992jz,
BraunMunzinger:1994xr, Cleymans:1996cd, Yen:1997rv,
BraunMunzinger:1999qy, Cleymans:1999st, BraunMunzinger:2001ip,
BraunMunzinger:2003zd, Karsch:2003zq, Tawfik:2004sw, Becattini:2005xt,
Andronic:2005yp, Andronic:2008gu,Begun:2012rf, Andronic:2012ut,
Tiwari:2011km, Fu:2013gga, Tawfik:2013eua, Garg:2013ata,
Bhattacharyya:2013oya, Bhattacharyya:2015zka, Kadam:2015xsa,
Kadam:2015fza, Kadam:2015dda, Albright:2014gva, Albright:2015uua,
Begun:2016cva}.  HRG model is not only successful in describing the
hadron yields in central heavy ion collisions from AGS up to RHIC
energies~\cite{BraunMunzinger:1994xr, Cleymans:1996cd,
BraunMunzinger:1999qy, Cleymans:1999st, BraunMunzinger:2001ip,
Becattini:2005xt, Andronic:2005yp, Andronic:2008gu} but also in
describing the bulk properties of hadronic matter in thermal and
chemical equilibrium \cite{Karsch:2003zq, Tawfik:2004sw,
Andronic:2012ut}.  The logarithm of the grand canonical partition function of a hadron
resonance gas can be written as \cite{Andronic:2012ut},
\begin {equation}
 \ln Z^{id}=\sum_i \ln Z_i^{id},
\end{equation}
where the sum is over all the hadrons. $id$ refers to ideal {\it i.e.},
non-interacting HRG. For particle species $i$,
\begin{equation}
 \ln Z_i^{id}=\pm \frac{Vg_i}{(2\pi)^3}\int d^3p
 \ln[1\pm\exp(-(E_i-\mu_i)/T)],
\end{equation}
where $V$ is the volume of the system, $g_i$ is the degeneracy factor,
$T$ is the temperature, $E_i$ is the single particle energy, $m_i$ is
the mass and $\mu_i=B_i\mu_B+S_i\mu_S+Q_i\mu_Q$ is the chemical
potential. In the last expression, $B_i,S_i,Q_i$ are respectively the
baryon number, strangeness and charge of the particle, $\mu^,s$ are
corresponding chemical potentials.  The upper and lower sign corresponds
to baryons and mesons respectively. We assume that the
hadronic matter is in thermal and chemical equilibrium therefore we have
ignored non-equilibrium phenomena like decays of particles along with
minimum biased jets and harmonisation.  We have ignored
the effect of parton fragmentation into hadrons which produces very
significant correlations at lower energies and in peripheral collisions
as high as 200 GeV \cite{Trainor:2010zv,Trainor:2012jv}. In addition,
at lower collision energies stopping becomes important which has not
been considered here. The partition function is the basic quantity from
which one can calculate various thermodynamic quantities of the thermal
system.  The number density $n_i$ of $i$ th particle is defined as,

\begin{equation}
 n_i =\frac{T}{V} \left(\frac{\partial \ln Z_i}
       {\partial\mu_i}\right)_{V,T} 
 =\frac{g_i}{{(2\pi)}^3} \int\frac{d^3p} {\exp[(E_i-\mu_i)/T]\pm1}.
\end{equation}

In case of heavy ion collision experiments, the parameters $T$ and
$\mu's$ of HRG model corresponds to those at chemical freeze-out which
depend on initial conditions of the collision.  The chemical potentials
$\mu_B, \mu_S$ and $\mu_Q$ are not independent, but related (on average)
to each other as well as to $T$ via the relations \cite{Alba:2014eba},
\begin{equation}
\label{eq:ns}
\sum_i n_i (T, \mu_B, \mu_S, \mu_Q) S_i=0,
\end{equation}
and
\begin{equation}
\label{eq:nbq}
 \sum_i n_i (T, \mu_B, \mu_S, \mu_Q) Q_i= r \sum_i n_i (T, \mu_B, \mu_S,
   \mu_Q) B_i,
\end{equation}
where $r$ is the ratio of net-charge to net-baryon number of the
colliding nuclei.  For Au + Au collisions $r = N_p /(N_p + N_n)=0.4$,
where $N_p$ and $N_n$ are respectively proton numbers and neutron
numbers of the colliding nuclei.  The Eq. \ref{eq:ns} is due to fact
that initially there is no net-strangeness in the colliding nuclei.  In
terms of transverse momentum $(p_T)$ and pseudo-rapidity ($\eta$), the
volume element $d^3p$ and the single particle energy $E_i$ can be
written as $d^3p=2\pi~p_T^2~ cosh~\eta ~dp_T ~d\eta$ and $E_i=\sqrt{(p_T
~cosh~\eta)^2+m_i^2}$, respectively.  Instead of pseudo-rapidity, one
can use rapidity ($y$) as well.  In that case $d^3p$ and $E_i$
respectively can be written as $d^3p=2\pi~p_T m_{Ti}~ cosh~y ~dp_T ~dy$
and $E_i=m_{Ti} cosh~y$, where $m_{Ti}=\sqrt{p_T^2+m_i^2}$.  The
experimental acceptances can be incorporated by considering the
appropriate integration ranges, either in $p_T$ and $\eta$ or in $p_T$
and $y$.

\subsection{Excluded volume corrections}

In ideal HRG model point like particles are considered. Although,
attractive interactions between hadrons are incorporated through the
presence of resonances, repulsive interactions are ignored in this
framework. This simple model has few parameters only.  Despite its
simplicity, this model successfully describes the bulk properties of the
system created in heavy ion collisions.  The repulsive interactions are
also needed, especially at very high temperature and/ or large baryon
chemical potential, to catch the basic qualitative features of strong
interactions where ideal gas assumption becomes inadequate.  In the
EVHRG model \cite{Hagedorn:1980kb, Rischke:1991ke, Cleymans:1992jz,
Yen:1997rv, Begun:2012rf, Andronic:2012ut, Fu:2013gga, Tawfik:2013eua,
Bhattacharyya:2013oya, Kadam:2015xsa, Kadam:2015fza, Kadam:2015dda,
Albright:2014gva, Albright:2015uua}, hadronic phase is modelled by a gas
of interacting hadrons, where the geometrical sizes of the
hadrons are explicitly incorporated as the excluded volume correction,
to approximate the short-range repulsive hadron-hadron interaction.

\section{\label{sec:Fluctuations} Fluctuations of conserved charges}

Derivatives of the $\ln Z$ with respect to corresponding chemical
potential define susceptibilities, which experimentally become
accessible through event-by-event analysis of fluctuations of conserved
quantities such as net-baryon number, net-charge, net-strangeness and
others.

The $n~th$ order susceptibility is defined as,
\begin{equation}\label{eq:chi}
 \chi^n_q=\frac{1}{V T^3}\frac{\partial^n {(\ln Z)}}{\partial
   {(\frac{\mu_q}{T})}^n}
\end{equation}
where $\mu_q$ is the chemical potential for conserved charge $q$. 

Experimentally net-charges $N_q$ ($= N_q^+-N_q^-$) are measured in a
finite acceptance on an event by event basis. The mean ($M_q$), variance
($\sigma_q^2$), skewness ($S_q$) and kurtosis ($\kappa_q$) of net-charge
distribution are related to the different order of susceptibilities by
the following relations:
\begin{equation}
M_q=\left\langle N_q \right \rangle = VT^3\chi_q^1,
\end{equation}
\begin{equation}
 \sigma_q^2=\left\langle(\delta{ N_q})^2\right\rangle=VT^3\chi_q^2,
\end{equation}
\begin{equation}
 S_q=\frac{\left\langle(\delta{ N_q})^3\right\rangle}{\sigma_q^3}=\frac{VT^3\chi_q^3}{(VT^3\chi_q^2)^{3/2}},
\end{equation}
\begin{equation}
 \kappa_q=\frac{\left\langle(\delta{ N_q})^4\right\rangle}{\sigma_q^4}-3=\frac{VT^3\chi_q^4}{(VT^3\chi_q^2)^2},
\end{equation}
where $\delta{ N_q} = N_q - \left\langle N_q \right\rangle$.  The mean,
variance, skewness and kurtosis are respectively estimations of the most
probable value, width, asymmetry and the peakedness of the distribution.
From the above equations, volume independent ratios can be defined by
the following relations:
\begin{subequations}
\label{allequations}
\begin{eqnarray}
&\sigma_q^2/M_q=C_2/C_1=\chi_q^2/\chi_q^1,\label{equationa}
\\
&S_q \sigma_q = C_3/C_2 = \chi_q^3/\chi_q^2,\label{equationb}
\\
&\kappa_q \sigma_q^2 = C_4/C_2 = \chi_q^4/\chi_q^2,\label{equationc}
\end{eqnarray}
\end{subequations}
where $C_n$ is the $n$ th order cumulants of the charge distribution.
The STAR collaboration has reported results of the above-mentioned
observables of net-proton and net-charge at different energies ranging
from $7.7$ GeV to $200$ GeV and at various centralities
\cite{Adamczyk:2013dal, Adamczyk:2014fia}.  The PHENIX collaboration has
also reported results of similar observables for net-charge
\cite{Adare:2015aqk}.  Non-monotonic variations of these ratios  with
beam energy ($\sqrt{s_{NN}}$) and also with centrality at a fixed
$\sqrt{s_{NN}}$ are believed to be good signatures of a phase transition
and a CEP.  These observables have also been studied in different models
\cite{Bhattacharyya:2013oya, Garg:2013ata, Alba:2014eba,
Bhattacharyya:2015zka, Albright:2015uua, Karsch:2015zna,
Ichihara:2015kba, Xu:2016jaz, Xu:2016qzd} and also in LQCD
\cite{Gupta:2011wh, Karsch:2011gg, Karsch:2015nqx, Bazavov:2015zja}.
Recently $S\sigma$ and $\kappa \sigma^2$ for charged pions have been
studied using non-equilibrium HRG model \cite{Begun:2016cva}.

\begin{figure*}[!htb]
\centering
 \includegraphics[width=\textwidth]{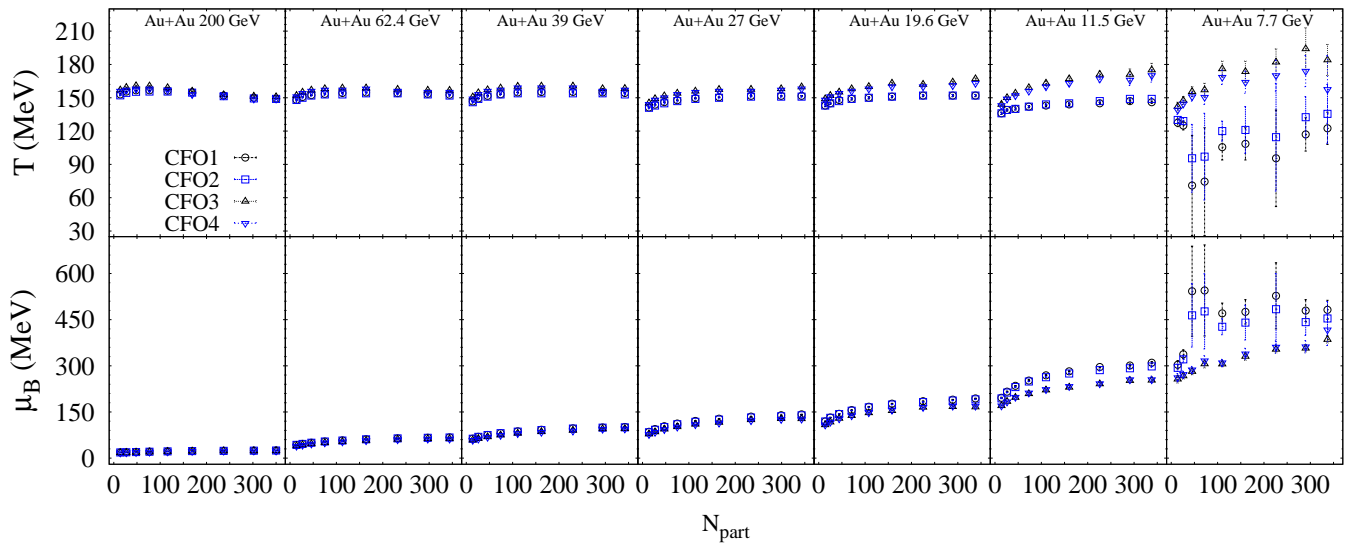}
 \caption{(Color online) Centrality (in terms of $N_{part}$) dependence
of chemical freeze-out temperatures and baryon chemical potentials for
Au + Au collisions at $\sqrt{s_{NN}}=~200,~62.4,~39,~27, ~19.6,~11.5$
and $7.7$ GeV. $\sqrt{s_{NN}}$ varies column wise. Four sets of chemical
freeze-out parameters (CFO) have been plotted.  For specifications of
different sets of parameters see the table \ref{tableLabel1}.}
 \label{T_mu_Npart}
\end{figure*}

\section{\label{sec:Results} Results}

In this paper, we have studied fluctuations of net-proton and net-charge
using HRG as well as its interacting version $i.e.$ EVHRG model.  In
Ref. \cite{Fu:2013gga, Bhattacharyya:2013oya}, it was shown that the
ratios of higher order cumulants are affected by the excluded volume
corrections.  Further, it was shown that experimental data of
$\sigma^2/M$ for net-proton as well as for net-charge in central Au + Au
collisions \cite{Bhattacharyya:2013oya} can be described quite well
using this model. Not only that, $S\sigma$, $\kappa\sigma^2$ can also be
described within experimental error for $\sqrt{s_{NN}} \ge 27$ GeV.
Therefore, it is very important to consider EVHRG model for the study of
fluctuations of conserved charges.  On the other hand ratio of cumulants
depend on acceptance cuts as well \cite{Garg:2013ata,
Bhattacharyya:2013oya,Karsch:2015zna}.  Therefore, in this work we have
used the HRG / EVHRG model with proper experimental acceptances.
For our present study we constrained the chemical
freeze-out temperature and chemical potentials using some of the
net-charge and net-proton measured cumulants and then predicted the
others in order to test the model. In all our calculations, we have
incorporated all the hadrons listed in the particle data book up to a
mass of 3 GeV \cite{Agashe:2014kda}.

\subsection{\label{sec:CFO} Centrality dependence of chemical freeze-out
parameters}

\begin{table}
\centering
\begin{tabular}{|c|c|c|}
\hline
Sets of parameters &Experimental data used&Model used\\
\hline
CFO1 &$(\sigma^2/M)_{np},(\sigma^2/M)_{nc}$&HRG \\
CFO2 &$(\sigma^2/M)_{np},(\sigma^2/M)_{nc}$ &EVHRG \\
CFO3 &$(\sigma^2/M)_{nc}$, $(S\sigma)_{np}, (S\sigma)_{nc}$ &HRG \\
CFO4 &$(\sigma^2/M)_{nc}$, $(S\sigma)_{np}, (S\sigma)_{nc}$ &EVHRG \\
\hline
\end{tabular}
\caption{Sets of chemical freeze-out parameters ($T, \mu's$). Subscripts
``np'' and ``nc'' correspond to net-proton and net-charge respectively.}
\label{tableLabel1}
\end{table}

The thermal fireball created due to heavy ion collision expands and
cools.  After some time inelastic collisions among the particles stop
and hence particle yields (or particle ratios) get fixed.  This stage is
called chemical freeze-out. From the experimental information about
particle yields or particle ratios, chemical freeze-out temperature and
baryon chemical potential can be estimated
\cite{Cleymans:2005xv,Xu:2001zj,Becattini:2005xt,Andronic:2005yp,
Andronic:2009jd, Karsch:2010ck,Chatterjee:2015fua}.  Chemical
freeze-­out parameters are reported to be independent of centrality
\cite{Kaneta:2004zr, Cleymans:2004pp}.  However, we wanted to revisit
the centrality dependence of chemical freeze­‐out parameters through the
study of higher order cumulants in Au + Au collisions.  Therefore, in
this paper, we estimate chemical freeze-out temperatures and
chemical potentials within HRG model, at different energies as well as
at different centralities using the experimentally measured ratios of
cumulants of net-proton and net-charge measured in Au + Au collisions by
STAR collaboration at RHIC \cite{Adamczyk:2013dal, Adamczyk:2014fia}.

Net-proton fluctuations were experimentally measured in the mid rapidity
($|y|<0.5$) and within transverse momentum $0.4 < p_T < 0.8$ GeV.  On
the other hand, net-charge fluctuations were measured in pseudo-rapidity
range $|\eta|<0.5$ and within transverse momentum range $0.2 < p_T <
2.0$ GeV (removing net-proton of $p_T<0.4$ GeV) \cite{Adamczyk:2014fia}.
Same acceptances have been used in the HRG / EVHRG model in the present
study.  We have considered hard core radii $0.3$ fm for all the hadrons
whenever EVHRG is used.  Four sets of chemical freeze-out parameters,
listed in table \ref{tableLabel1}, have been used in order to
describe $\sigma^2/M$, $S\sigma$ and $\kappa\sigma^2$ of net proton and
net-charge.

Here we would like to discuss the modus operandi for the
estimation of parameter sets listed in table \ref{tableLabel1}. We have
three experimental cumulant ratios $\sigma^2/M$, $S\sigma$ and
$\kappa\sigma^2$ for net charge and net proton. It should be noted that
$\sigma^2/M$ has smaller experimental errors compared to $S\sigma$ and
$\kappa\sigma^2$.  Not only that, experimental errors are smaller for
net-proton fluctuations compared to net-charge data. In order to
evaluate the chemical freeze-out parameters from these observables at a
particular $\sqrt{s_{NN}}$ and centrality, we use $\chi^2$ minimisation
technique where $\chi^2$ is defined as,
\begin{equation}\label{chi2_min}
\chi^2=1/N \sum_{i=1}^{N}
\frac{(R_i^{expt}-R_i^{model})^2}{\sigma_i^2},
\end{equation}
where $N$ is the number of observables, $R_i^{model}$ is the $i ~th$
observable with $R_i^{expt}$ and $\sigma_i$ being its experimental
values and errors respectively (statistical error has been used here).
Error bars in the evaluated freeze-out parameters correspond to
$\chi^2=\chi^2_{min}+1$. We have taken care of the conservation laws
Eqs. \ref{eq:ns} and \ref{eq:nbq} in the evaluation of chemical
freeze-out parameters.

First we obtained freeze-out parameters using only
lower order cumulant ratios $\sigma^2/M$ of net-charge and net-proton.
For this we have the two sets CFO1 (HRG) and CFO2 (EVHRG). Then we
wanted to check if the freeze-out parameters estimated including the
higher order cumulant $S\sigma$ for net-charge and net-proton agrees
with the above set.  We found that the extremely high precision of
experimental data for $\sigma^2/M$ of net-proton completely biases the
$\chi^2$ minimisation to agree with the earlier set. So finally we used
$\sigma^2/M$ and $S\sigma$ of net-charge and $S\sigma$ of net-proton to
extract the second set of parameters CFO3 (HRG) and CFO4 (EVHRG). Note
that if equilibration is complete then any combination of observables
should reproduce mutually agreeable set of fitting parameters.

\begin{figure*}[!htb]
\centering
 \includegraphics[width=\textwidth]{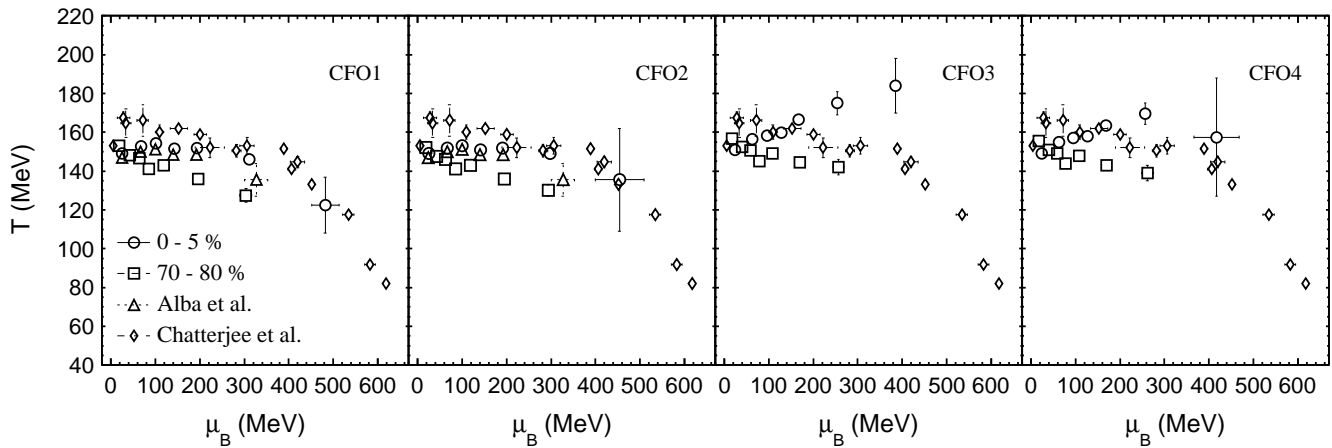}
 \caption{Chemical freeze-out parameters in ($T, \mu_B$) plane of the
QCD phase diagram. We plot chemical freeze-out parameters for most
central ($0- 5 \%$) as well as for most peripheral ($70- 80 \%$)
collisions for $7.7$ GeV $\le \sqrt{s_{NN}} \le 200$ GeV.  We compare
our results of CFO1 and CFO2 with chemical freeze-out parameters for $0-
5 \%$ centralities given in the Ref. Alba et al. \cite{Alba:2014eba} for
$11.5$ GeV $\le \sqrt{s_{NN}} \le 200$ GeV. Chemical freeze-out
parameters given in the Ref. Chatterjee et al. \cite{Chatterjee:2015fua}
have also been plotted.}
\label{T_muB}
\end{figure*}

Figure \ref{T_mu_Npart} shows centrality dependence of chemical
freeze-out $T, \mu_B$ at different $\sqrt{s_{NN}}$ from $7.7$ GeV up to
$200$ GeV.  Four sets of chemical freeze-out parameters (CFO1 - CFO4)
have been plotted.  The average number of participant ($N_{part}$) is
maximum for most central ($0 - 5 \%$) collision whereas it is minimum
for most peripheral ($70 - 80 \%$) collision.  In this figure
$\sqrt{s_{NN}}$ decreases column wise.  The leftmost column of Fig.
\ref{T_mu_Npart} corresponds to the highest beam energy $i.e.
\sqrt{s_{NN}}= 200$ GeV whereas the right most column corresponds to
$\sqrt{s_{NN}}= 7.7$ GeV.

One would expect that with higher $\sqrt{s_{NN}}$ the particle
production would be higher and give rise to a high freeze-out
temperature. On the other hand for low $\sqrt{s_{NN}}$ particle
production would be less and the collision participants would also
contribute actively to the system properties (due to baryon stopping).
Thus observed temperature may be low but baryon chemical potential may
be large. Note that for complete equilibration at freeze-out, all
evolutionary history of the system will be erased. This will be
reflected in the agreement of thermodynamic parameters fitted from all
possible experimental observables. On the other hand, for incomplete
equilibration certain discrepancies among the thermodynamic parameters
fitted from different observables may arise.  On top of that the
presence of jets, hadronic decays, as well as
interactions among the hadrons beyond those considered through the
excluded volume effects, may also show deviation of the system from that
expected from the HRG picture used to model the system.

It can be seen from Fig. \ref{T_mu_Npart} that chemical freeze-out
temperatures of CFO1 and CFO3 (or CFO2 and CFO4) are almost the same for
$\sqrt{s_{NN}}\ge 27$ GeV. There are significant separations between
these two sets of parameters for $\sqrt{s_{NN}}< 27$ GeV, and these
separations increase towards central collisions for a fixed
$\sqrt{s_{NN}}$. Not only that, these separations increase with the
decrease of $\sqrt{s_{NN}}$. Overall the spread in temperature for the
whole range of $\sqrt{s_{NN}}$ and centrality is within 140-180 MeV.  On
the other hand the magnitudes of chemical freeze-out baryon chemical
potentials ($\mu_B$) increase with decrease of $\sqrt{s_{NN}}$ as well
as increase in centrality by about two orders of magnitude.  The
occurrence of high net-baryon density is expected when the participant
nucleons are deposited in the collision region. More or less similar
behaviour of $\mu_B$ is reported in Ref.
\cite{Kumar:2012np,Das:2014oca,Yu:2014epa} where chemical freeze-out
parameters are extracted analysing particle yields measured
experimentally \cite{Kumar:2012np,Das:2014oca} or generated by the event
generator \cite{Yu:2014epa}.  The separation of the parameters obtained
from CFO1 and CFO3 (or CFO2 and CFO4) are also observed here, but in the
opposite direction. The lower order cumulants thus seems to equilibrate
with lower temperature and higher density than the higher order
cumulants.

The conclusion that one can draw from this observation is that the
system formed in the heavy ion collision has not completely equilibrated
if we consider only the HRG model to describe it. However it is possible
that HRG picture, if suitably modified, may lead to the scenario as
found in Fig.~\ref{T_mu_Npart}. Here the question is whether there are
any possibilities such that one can find a multicomponent system with
different equilibrium parameters that can systematically explain the
observed discrepancy for the fitted parameters.

At this point it is tempting to propose a possible scenario that may
give rise to such an agreement of thermodynamic parameters for higher
$\sqrt{s_{NN}}$ and deviations found for lower $\sqrt{s_{NN}}$.
We first assume that in the region of lower values of
$\sqrt{s_{NN}}$ the effects coming from the jets are quite small, and
hence are not responsible for this deviation.
Now if the system has thermalised near or above the phase
transition region and then evolved down to the hadronic phase then one
can qualitatively describe the situation as follows. For a cross-over
region the system undergoes rapid changes from partonic to hadronic
phase, but all the components can still follow a given equilibrium
condition at all times.  This is expected to happen for large
$\sqrt{s_{NN}}$. However near the critical point, correlation length $\xi$
would tend to infinity and there would be a large enhancement in the
fluctuations.  In a realistic situation, as in heavy ion collisions,
dynamical variables are functions of time.  As the system moves towards
the critical region, relaxation time increases and at some point the
system may expand too fast to maintain thermodynamic equilibrium. So the
correlation length gets constrained due to this critical slowing down
\cite{Ma_1976} and becomes frozen at some time.  But the system
expansion continues further. This situation may lead to the difference
in the information carried by the different order of cumulants.  More
specifically, second, third and fourth order cumulants of multiplicities
are related to the correlation length by the relations
$\left\langle(\delta{N})^2\right\rangle \sim \xi^2$,
$\left\langle(\delta{N})^3\right\rangle \sim \xi^{4.5}$ and
$\left\langle(\delta{N})^4\right\rangle \sim \xi^7$ respectively
\cite{Stephanov:2008qz}.  This, in turn, implies that for higher order
cumulants their relaxation time to the equilibrium values may be
considerably larger compared to those of lower order cumulants. So in
the final spectrum, higher order cumulants are expected to carry the
information of the system farther from equilibrium compared to lower
cumulants. For example, compared to lower moments, the temperature
evaluated using the higher moments may be larger as being away from
equilibrium system is hotter. This is what we observe for lower
$\sqrt{s_{NN}}$ i.e. temperatures of CFO3 / CFO4, where third order
fluctuations are involved, are larger compared to that of  CFO1 / CFO2
for $\sqrt{s_{NN}}< 27$ GeV, the corresponding chemical potential being
smaller than that of CFO1/CFO2.  Incidentally this is the range of
temperature and baryon chemical potential where close to which
the critical end point is expected to lie.  Availability of higher
moment data with much better statistics is extremely essential for
further constraining this picture. We however emphasise that this is
only a plausibility argument for effects of a CEP to modify the simple
HRG parameters with different cumulants. A systematic study of various
other dynamical effects would be required to ascertain how far this
picture is valid \cite{Luo:2014tga,Netrakanti:2014mta}.
Another important caveat is that the contributions due to purely
statistical fluctuations in the cumulants reported by the STAR
experiment are not subtracted from the variances and rely on models
for $S\sigma$ and $\kappa\sigma^2$. Therefore, the sensitivity of the
reported cumulants to dynamical effects is ambiguous. In fact it is
even difficult to ascertain whether the statistical fluctuations in
the data may overwhelm the critical fluctuations or not.

Figure \ref{T_muB} shows chemical freeze-out parameters for 0-5$\%$ and
$70-80 \%$ in the ($T, \mu_B$) plane.  In this figure, we also compare
our results of chemical freeze-out parameters with previous works
\cite{Alba:2014eba, Chatterjee:2015fua}.  In \cite{Alba:2014eba},
chemical freeze-out parameters for most central collisions were
estimated using the experimental data of $\sigma^2/M$ of
net-proton and net-charge. In their model they considered effect of the
resonance decays, experimental acceptances and randomisation of the
isospin of nucleons in the hadronic phase. They excluded chemical
freeze-out parameters for $\sqrt{s_{NN}} = 7.7$ GeV.  It can be seen
that our chemical freeze-out parameters of CFO1 / CFO2 for $0 - 5 \%$
centrality are very close to that of \cite{Alba:2014eba}  for
$\sqrt{s_{NN}} = 19.6 - 200$ GeV. Moreover, the agreement is slightly
better for CFO2.  Interesting point is that, with decrease of
$\sqrt{s_{NN}}$ from $\sqrt{s_{NN}} = 200$ GeV, chemical freeze-out $T$
increases up to $\sqrt{s_{NN}} = 39$ GeV then it decreases at
$\sqrt{s_{NN}} = 27$ GeV and becomes almost flat up to $\sqrt{s_{NN}} =
19.6$ GeV and then again decreases.  In contrast, the chemical
freeze-out $\mu_B$ increases with decrease of $\sqrt{s_{NN}}$ in the
whole range of $\sqrt{s_{NN}}$.  This behaviour of chemical freeze-out
$T$ is in contradiction to what has been reported  in the Refs.
\cite{Cleymans:2005xv,Xu:2001zj,Becattini:2005xt,Andronic:2005yp,
Andronic:2009jd,Karsch:2010ck,Tiwari:2011km,Chatterjee:2015fua} where
chemical freeze-out parameters have been extracted from
particle multiplicities.  We plot chemical freeze-out $T$ and $\mu_B$ of
Ref.  \cite{Chatterjee:2015fua} for comparison.  Refs.
\cite{Cleymans:2005xv,Xu:2001zj,Becattini:2005xt,Andronic:2005yp,
Andronic:2009jd,Karsch:2010ck,Tiwari:2011km,Chatterjee:2015fua} showed
that chemical freeze-out $T$ rapidly increases with the increase of
$\sqrt{s_{NN}}$ in SIS-AGS-SPS energy range and then saturates at top
RHIC energy.  However, behaviour of chemical freeze-out $\mu_B$ reported
in these references were similar.  In Fig. \ref{T_muB} we also show
chemical freeze-out $T, \mu_B$ of CFO3/ CFO4.  For both CFO3 and CFO4,
chemical freeze-out $T$ increases with decrease of $\sqrt{s_{NN}}$ for
$11.5$ GeV $\le \sqrt{s_{NN}} \le 200$ GeV and the temperatures are
larger compared to that of CFO1 / CFO2.  Although the fast rise of
$\mu_B$ for higher moments as found in HRG, seems to have slowed down
for EVHRG as seen in the figure for CFO4., all the chemical freeze-out
parameters are within certain band in the ($T, \mu_B$) plane. Recently,
the possibility of larger chemical freeze-out temperature is indicated
also at LHC energy \cite{Vovchenko:2015cbk}. 

\begin{figure*}[!htb]
\centering
 \includegraphics[width=\textwidth]{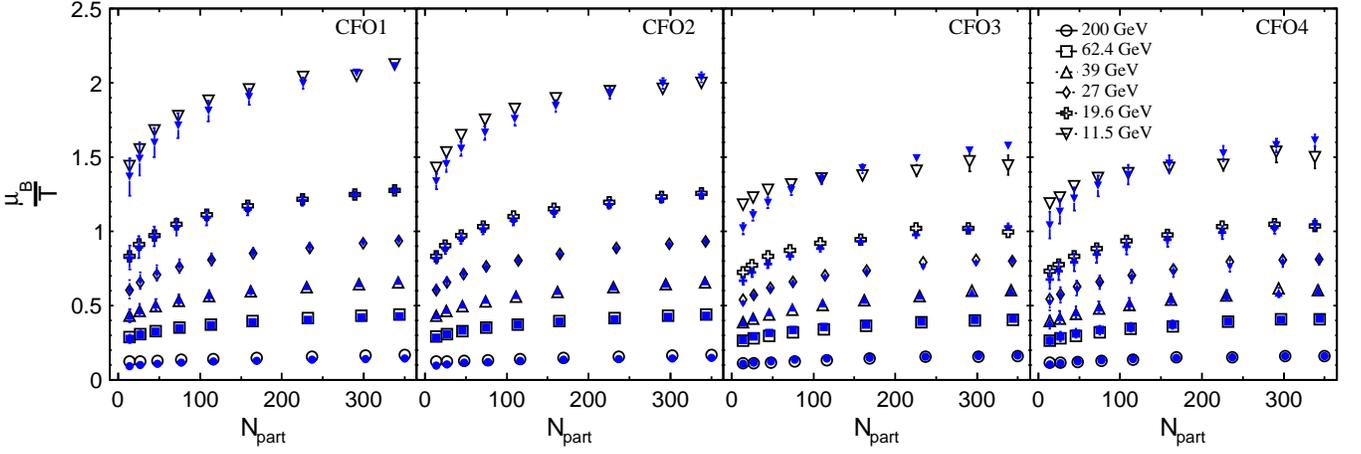}
 \caption{(Color online) Variation of $\mu_B/T$ with $N_{part}$ (Black
points). Blue solid points correspond to the $\mu_B/T$ according to the
Eq. \ref{fitting_eq}.}
 \label{fitting_muB_T}
\end{figure*}
\begin{figure*}[!htb]
\centering
\includegraphics[width=\textwidth]{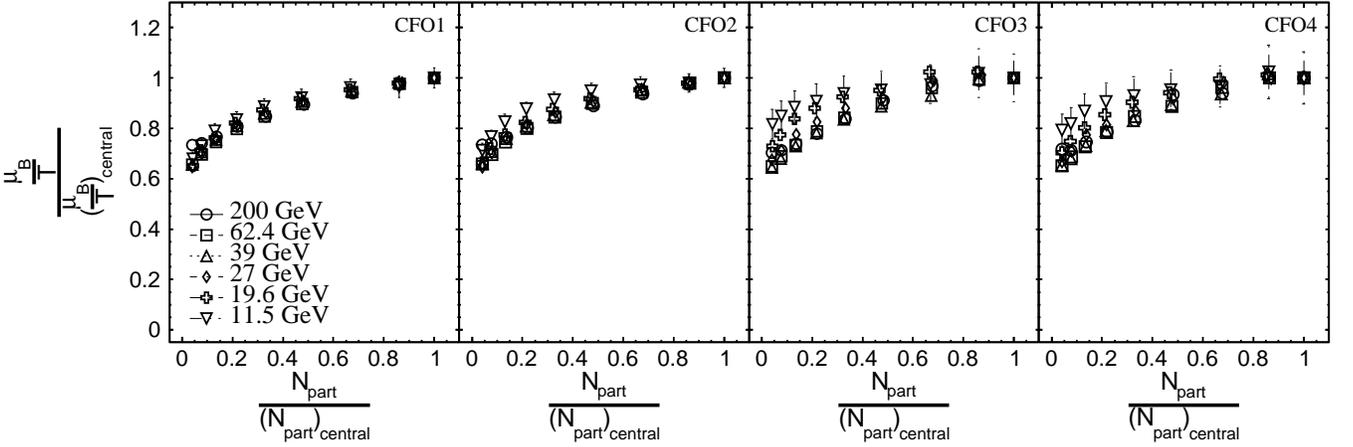}
 \caption{Scaling behaviour of $\mu_B/T$ with centrality.  On the
horizontal axis, $N_{part}$ is normalised with that at the most central
collision and similarly in the vertical axis $\mu_B/T$ is normalised to
that of the most central collision.  Therefore, in the horizontal axis,
the maximum value, which is equals to $1$, corresponds to the most
central collision ($0-5 \%$) and the minimum value corresponds to the
most peripheral collision ($70-80 \%$).}
 \label{muB_T_Npart}
\end{figure*}

\subsection{Scaling behaviour of \texorpdfstring{$\mu_B/T$}{mB/T}}
\label{sec:Scaling}

The Fig. \ref{fitting_muB_T} shows variation of $\mu_B/T$ with
$N_{part}$.  $\mu_B/T$ increases with increase of $N_{part}$ for all
energies for all four parameter sets.  Moreover, $\mu_B/T$ increases
with decrease of $\sqrt{s_{NN}}$.  It can also be seen that $\mu_B/T$
for CFO1 / CFO2 are larger compared to CFO3 / CFO4 and differences
between the parameters of CFO1 and CFO3 (or CFO2 and CFO4), as shown in
Fig. \ref{T_muB}, increase when the value of $\mu_B/T$ is close to or
greater than unity.  In order to separate the effects of $N_{part}$ and
$\sqrt{s_{NN}}$,  $\mu_B/T$ can be expressed by the relation,
\begin{equation}\label{fitting_eq}
 \mu_B/T(\sqrt{s_{NN}},N_{part})= p(0) (N_{part})^{1/p(1)}
\sqrt{s_{NN}}^{p(2)},
\end{equation}
where $p(0), p(1)$ and $p(2)$ are three parameters. In this equation,
first part depends only on $N_{part}$ while second part depends only on
$\sqrt{s_{NN}}$.  For fitting purpose we have simultaneously used
$\mu_B/T$ of $\sqrt{s_{NN}}= 19.6$ GeV to $62.4$ GeV for CFO1 / CFO2 and
$\sqrt{s_{NN}}= 27$ GeV to $200$ GeV for CFO3 /CFO4 for which $\chi^2$
per degree of freedom (ndf) is minimum.  All the fitting parameters are
listed in the table \ref{tableLabel2}. The quality of fitting is quite
good as can be seen from the figure.  The fitted parameters are then
used to estimate $\mu_B/T$ for remaining energies. 
It can be seen that,
for the sets CFO1 / CFO2, the $\mu_B/T$ from Eq. \ref{fitting_eq}
slightly underestimates the extracted $\mu_B/T$ for
$\sqrt{s_{NN}}= 11.5$ GeV and $200$ GeV. On the other hand, for CFO3/ CFO4,  $\mu_B/T$
evaluated using Eq.  \ref{fitting_eq} slightly underestimates the
extracted $\mu_B/T$ for peripheral collisions of
$\sqrt{s_{NN}}= 11.5$ GeV and slightly overestimates towards central
collisions.

\begin{table*}
\centering
\begin{tabular}{|c|c|c|c|c|c|}
\hline
CFO &Using $\frac{\mu_B}{T}$ of&$p(0)$&$p(1)$&$p(2)$&$\frac{\chi^2}{ndf}$\\
& $\sqrt{s_{NN}} (GeV)$ &&&&\\
\hline

CFO1& 19.6-62.4 &$9.59 \pm 1.33$ & $7.30 \pm 1.60$ & $-0.95 \pm 0.02$ &0.13\\

CFO2& 19.6-62.4 & $8.95 \pm 0.28 $& $7.56 \pm 0.40$ & $-0.92 \pm 0.01$ & 0.17\\

CFO3& 27-200 &$4.9 \pm 0.14$ &$7.28 \pm 0.34$  & $-0.79 \pm 0.01$ & 0.53\\
CFO4& 27-200 &$5.23 \pm 0.44$ &$7.29 \pm 0.97$  & $-0.81 \pm 0.02$ & 0.11\\
\hline
\end{tabular}
\caption{Parameters of the fitting function $\mu_B/T=p(0)
(N_{part})^{1/p(1)} \sqrt{s_{NN}}^{p(2)}$.  Since $\frac{\mu_B}{T}$ is
dimensionless, the dimension of p(0) is equals to GeV$^{-(p(2))}$.}
\label{tableLabel2}
\end{table*}


In the Fig. \ref{muB_T_Npart}, we have explored the  scaling behaviour
of $(\mu_B/T)/(\mu_B/T)_{central}$ with $N_{part}/(N_{part})_{central}$.
Quantities in both the axes have been normalised to the corresponding
values in the most central collisions.  As a result, the maximum value
in the horizontal axis (equals to unity) corresponds to most-central
collision ($0-5 \%$) and its value decreases towards most peripheral
collisions ($70-80 \%$).  It can be seen that  $\frac{\mu_B}{T}/
(\frac{\mu_B}{T})_{central}$ increases with increase in
$N_{part}/(N_{part})_{central}$ for all the $\sqrt{s_{NN}}$ from $200$
GeV down to $11.5$ GeV.  For most peripheral collisions,
$\frac{\mu_B}{T}/ (\frac{\mu_B}{T})_{central}$ become within $65- 85 \%$
of that of central collisions.  $\frac{\mu_B}{T}/
(\frac{\mu_B}{T})_{central}$ for all the $\sqrt{s_{NN}}$ seem to scale
well for the parameter sets CFO1/CFO2.  On the other hand, scaling is
found to be violated at lower $\sqrt{s_{NN}}$ for CFO3 / CFO4.  The
violation is large for $\sqrt{s_{NN}}=11.5$ GeV and small for
$\sqrt{s_{NN}}=19.6$ GeV. This violation of scaling
may again be interpreted as due to possible critical behaviour at lower
$\sqrt{s_{NN}}$, apart from being caused by other dynamical effects, as
already discussed earlier. It can be noted that here separations are
observed towards lower values of the horizontal axis because we have
normalised both the axes with the corresponding values in the most
central collisions.  However, normalisation of the axes with
corresponding values from most peripheral collisions would have
resulted in separations towards higher values of the horizontal axis.
In the above discussion $\sqrt{s_{NN}}= 7.7$ GeV have been excluded due
to large error bars.

The presence of scaling is a direct consequence of the fact that one can
separate the dependence of $\mu_B/T$ on $\sqrt{s_{NN}}$ and $N_{part}$
as given by Eq. \ref{fitting_eq}. By construction,
$(\mu_B/T)/(\mu_B/T)_{central}$ becomes independent of $\sqrt{s_{NN}}$.
However, since the fitting of $\mu_B/T$ is not perfect (Fig.
\ref{fitting_muB_T}) for all the $\sqrt{s_{NN}}$, scaling as shown in
Fig. \ref{muB_T_Npart} is also not exact.

\subsection{\label{sec:Comparison} Comparison with experimental data}

\begin{figure*}[!htb]
\centering
 \includegraphics[width=\textwidth]
{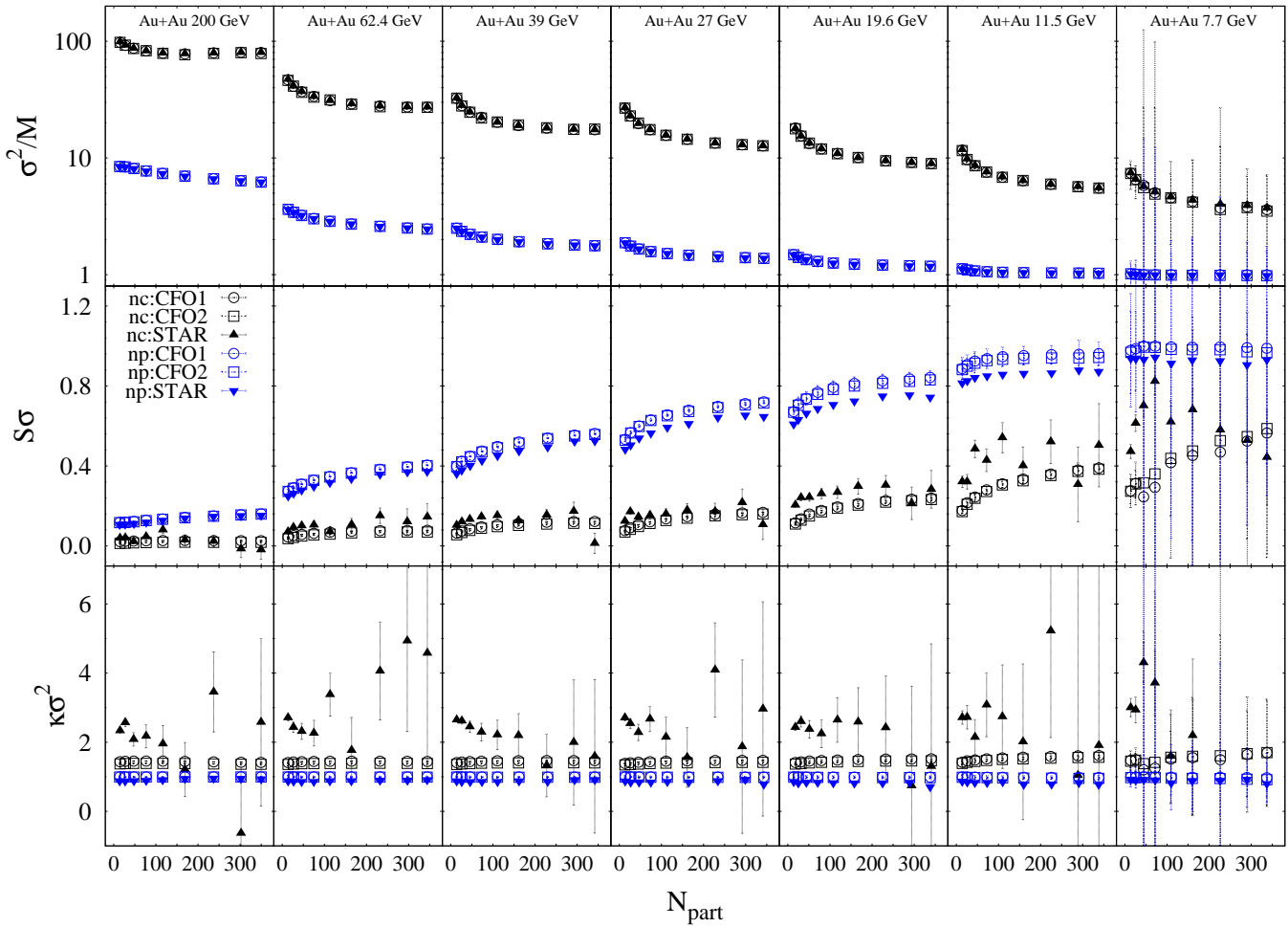}
\caption{(Color online) Centrality dependence of ratios of cumulants of
net-proton and net-charge. ``nc'' and ``np'' correspond to net-proton
and net-charge respectively. Experimental data of fluctuations measured
in Au + Au collisions by STAR collaboration is taken from
\cite{Adamczyk:2013dal, Adamczyk:2014fia}. Blue and black points have
been used for net-proton and net-charge respectively.}
 \label{moments_products_np_nc_2}
\end{figure*}

\begin{figure*}[!thb]
\centering
 \includegraphics[width=\textwidth]
{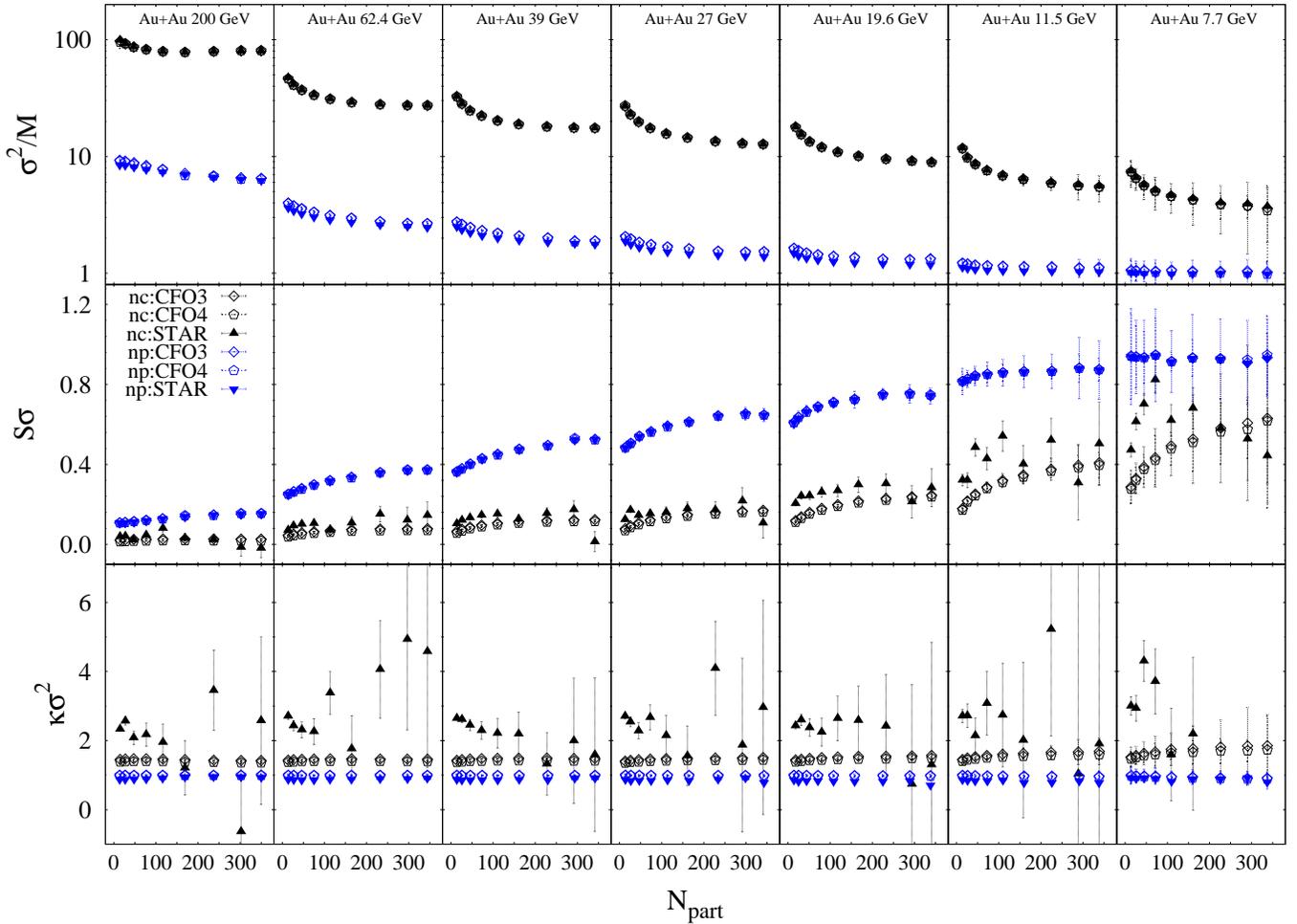}
  \caption{(Color online) Same as Fig. \ref{moments_products_np_nc_2}
but for parameter sets CFO3 and CFO4.}
\label{moments_products_np_nc_3}
\end{figure*}

Here we have used the extracted freeze-out parameters CFO1
and CFO2 to calculate the $\sigma^2/M$, $S\sigma$ and $\kappa\sigma^2$
of net-proton and net-charge using HRG and EVHRG model respectively.
Note that for these two sets, chemical freeze-out parameters were
estimated from experimental data of $\sigma^2/M$ only.  We
have compared our results with experimental data of fluctuations
measured in Au + Au collisions by STAR collaboration
\cite{Adamczyk:2013dal, Adamczyk:2014fia}.  As mentioned earlier the
experimental acceptances have been incorporated in our model calculation
as well.

In the top row of Fig. \ref{moments_products_np_nc_2}, we have shown
centrality dependence of $\sigma^2/M$ of net-proton and net-charge at
different beam energies. In this figure $\sqrt{s_{NN}}$ varies column
wise. Blue and black points have been used for net-proton and
net-charge respectively in this figure.  For both net-proton and
net-charge $\sigma^2/M$ decreases with increase of $N_{part}$ for all
$\sqrt{s_{NN}}$. It can be seen that the ratios of lowest order
susceptibilities ($i.e.$ $\sigma^2/M$) of net-proton and net-charge can
be reproduced quite well using the CFO1 and CFO2 parameters in our
model.

We now evaluate the higher order susceptibility ratios using these two
parameter sets. The middle row of Fig. \ref{moments_products_np_nc_2}
shows centrality dependence of $S\sigma$ of net-proton and net-charge.
The quantity $S\sigma$ for both net-proton and net-charge increases with
increasing $N_{part}$ for  $\sqrt{s_{NN}} \ge 11.5$ GeV. Experimental
data of $S\sigma$ of net-proton also shows a similar trend.

The experimental data of $S\sigma$ of net-proton matches within the
error bar for $\sqrt{s_{NN}} > 27$ GeV.  However, its value is
overestimated at lower energies ($\sqrt{s_{NN}}\le 27$ GeV).  On the
other hand $S\sigma$ of net-charge calculated in the HRG / EVHRG model
are close to or within the error bars of experimental data for all
centralities at $\sqrt{s_{NN}} \ge 27$ GeV and for $\sqrt{s_{NN}} < 27$
GeV most central data matches within error bars.  In general $S\sigma$
of net-charge shows a monotonic behaviour and differs considerably from
experimental data.   

In the bottom row of Fig. \ref{moments_products_np_nc_2} we have shown
centrality dependence of $\kappa\sigma^2$ of net-proton and net-charge.
The $\kappa\sigma^2$ for net-proton calculated in our model using CFO1 /
CFO2 are almost independent of centrality for $\sqrt{s_{NN}} \ge 19.6$
GeV and below that energy it decreases slightly with increase of
$N_{part}$.  For all $\sqrt{s_{NN}}$, $\kappa\sigma^2$ of net-proton
calculated in the HRG / EVHRG model are within the error bars or very
close to the experimental data.  The experimental data of
$\kappa\sigma^2$ for net-charge matches within the error bar with the
HRG / EVHRG model results as we go towards central collisions for
$\sqrt{s_{NN}} \ge 11.5$ GeV.

Therefore we see that the HRG prediction of higher order cumulants
calculated from the estimated freeze-out thermodynamic parameters for the
lower order cumulants do not match the experimental data in general.
This implies that the equilibration at freeze-out is not quite
comprehensive vis-a-vis the HRG model. We can further check what happens
when we consider one more higher order cumulant as discussed below.

Fig. \ref{moments_products_np_nc_3} correspond to similar plot, as in
Fig. \ref{moments_products_np_nc_2}, for parameter sets CFO3 and CFO4.
While $\sigma^2/M$ for both net-proton and net-charge show consistency
when compared to the experimental data as shown in the top row, the use
of CFO3 / CFO4 in the HRG / EVHRG model, show clear improvement in
agreement with experimental data of $S\sigma$ for net-proton at all
$\sqrt{s_{NN}}$ as shown in the middle row. On the other hand, there is
almost no change in the results for $S\sigma$ of net-charge. Once again
we find that the prediction for $\kappa\sigma^2$ for net-proton agrees
well with experimental data whereas that for net-charge does not. This
further reaffirms that the matter formed in heavy ion collision system
does not conform to a system of completely equilibrated hadron gas.

\section{\label{sec:Discussion} Discussion and Conclusion}

We have extracted the chemical freeze-out
parameters by fitting the experimental data of cumulants of net-proton
and net-charge measured by STAR collaboration using both HRG and EVHRG
model. We have incorporated the proper experimental acceptances in our
calculation.  However, the dynamical effects such as particle decay,
minimum biased jet, baryon stopping are not considered in the
present study. The experimental data of $\sigma^2/M$ of both net-proton
and net-charge have been used to estimate chemical freeze-out
parameters CFO1 / CFO2.  On the other hand, parameters CFO3 / CFO4 have
been estimated using the experimental data of $\sigma^2/M$ of
net-charge and $S\sigma$ of both net-proton and net-charge. For CFO1 and
CFO3, HRG model has been used, whereas for CFO2 and CFO4, EVHRG has been
used. 

The chemical freeze-out parameters evaluated using lower order cumulants
(CFO1/CFO2) starts deviating from the one obtained using higher order
cumulants (CFO3/CFO4) around $\sqrt{s_{NN}} = 19.6$ GeV 
as one goes from $\sqrt{s_{NN}} = 200$ GeV towards lower energies.
Among other possibilities, transition of the system close to
the critical region may contribute to the requirement of multiple
parametrisation in HRG for various orders of cumulants. In case of
lower energies one need to take into account the baryon stopping as
well. In these regions of low energies HRG and EVHRG starts deviating
from each other as well due to the effect of repulsive interaction in
EVHRG.

We observe that the effect of centrality and beam energy in $\mu_B/T
(\sqrt{s_{NN}}, N_{part})$ can be separated. This separation leads to a
scaling of $(\mu_B/T)/(\mu_B/T)_{central}$ with
$N_{part}/(N_{part})_{central}$. Though the scaling is very good for
CFO1 / CFO2, a deviation is observed for CFO3 / CFO4 especially in the
region  $\sqrt{s_{NN}} \le 19.6$ GeV.  The study of such scaling
behaviour will be useful to search for CEP which is the main goal of the
ongoing STAR experiment and the future CBM experiment.

Experimental data of lowest order susceptibilities ($i.e.,$
$\sigma^2/M$) of net-proton and net-charge can be reproduced quite well
using the CFO1 and CFO2 in the HRG / EVHRG model. The experimental data
of $S\sigma$ of net-proton match within the error bar for $\sqrt{s_{NN}}
> 27$ GeV for these two sets of parameters.  However, it is
overestimated at lower beam energies ($\sqrt{s_{NN}}\le 27$ GeV).  On
the other hand $S\sigma$ of net-charge calculated in the HRG / EVHRG
model using CFO1 / CFO2 are close to or within the error bars for
$\sqrt{s_{NN}} \ge 27$ GeV and they are within the error bars for more
central data at lower $\sqrt{s_{NN}}$.  For all $\sqrt{s_{NN}}$,
$\kappa\sigma^2$ of net-proton calculated in the HRG / EVHRG model using
CFO1 / CFO2 are within the error bars or very close to the experimental
data. The experimental data of $\kappa\sigma^2$ for net-charge
matches within the error bar with the HRG / EVHRG model results
calculated using CFO1 / CFO2 as we move towards central collisions for
$\sqrt{s_{NN}} \ge 11.5$ GeV, but underestimate the data for peripheral
collisions.  This points to the incomplete equilibrium distribution of
the particles observed in data.  On the other hand experimental data of
$S\sigma$ of net-proton can be described well at all $\sqrt{s_{NN}}$ in
the HRG / EVHRG model using CFO3 / CFO4.  In addition, both the
parameter sets give satisfactory description of $\sigma^2/M$ of
net-proton and net-charge.  However the $\kappa\sigma^2$ for both
net-proton and net-charge calculated in the HRG / EVHRG model using CFO3
/ CFO4 are similar to those calculated using CFO1 / CFO2. In this set
again we found the signature of incomplete equilibration of the system
formed in heavy ion collision experiments.

Thus we conclude that the freeze-out parameters, which can
describe lower order cumulant ratios very well in all energies and
centralities, can't describe the higher order cumulant ratios equally
well. It is difficult to pin-point all the reasons for such disagreement
unless all the dynamical effects are accounted for. Looking at the
systematic deviation of the thermodynamic parameters we could only
present a plausibility argument for the system passing near a critical
region. Precise experimental data of $\kappa\sigma^2$ along with few
more $\sqrt{s_{NN}}$ around $19.6$ GeV will be extremely useful for
further investigation in this direction.

\section*{Acknowledgements}
This work is funded by CSIR, UGC, and DST of the Government of India. We
acknowledge the STAR collaboration for the experimental data.  SS thanks
Sabita Das for providing chemical freeze-out parameters of Ref.
\cite{Chatterjee:2015fua}. We would like to thank Bedangadas Mohanty for
useful discussions.

\bibliography{RefFile}

\end{document}